\documentstyle{article}

\def\be{\begin{equation}}
\def\ee{\end{equation}}
\def\bea{\begin{eqnarray}}
\def\eea{\end{eqnarray}}

\begin{document}

%\preprint{astro-ph/yymmddd}
%\draft 

%
% Remove this and closure after abstract, plus preprint number,
% in electronic submission
%
 
%
%\renewcommand{\topfraction}{0.99}
%\renewcommand{\bottomfraction}{0.99}
%@twocolumnfalse\endcsname
  
%\title
\begin{center}

\vskip 1cm
{\Large {\bf Cosmological dynamics of the tachyon with an inverse power-law potential
}}
       
\vskip 1cm

{\large L. Raul W. Abramo $^\dagger$, Fabio Finelli $^\ddagger$ 
\footnote{Also 
supported by INFN, Sezione di
Bologna, via Irnerio 46 -- I-40126 Bologna -- Italy}} \\
{~\\$^\dagger$ Instituto de F\'{\i}sica, Universidade de S\~ao Paulo \\
CP 66318, 05315-970 S\~ao Paulo, Brazil;
%~\\$^\ddagger$ Department of Astronomy, University of Bologna, Via
%Ranzani 1, I-40127 Bologna, Italy;
~\\$^\ddagger$ IASF/CNR, Istituto di Astrofisica Spaziale e
Fisica Cosmica \\ Sezione di Bologna \\ Consiglio Nazionale delle
Ricerche \\
via Gobetti, 101 -- I-40129
Bologna -- Italy
%/Bologna, Via Gobetti 101, I-40129, Bologna, Italy
}
%\date{\today} 
%\maketitle

%\begin{abstract} 
%\end{abstract}

%\pacs{PACS numbers: 98.80.-k, 98.80.Cq}]
\end{center}

\vskip 1cm

\abstract{
\noindent 
We investigate tachyon dynamics with an inverse power-law potential 
$V(\varphi) \propto \varphi^{-\alpha}$. 
We find global attractors of the dynamics leading to a dust 
behaviour for $\alpha > 2$ and to an
accelerating universe for $0 < \alpha \le 2$. We study linear 
cosmological perturbations and we show that metric
fluctuations are constant on large scales in both cases.
In presence of an additional perfect fluid, the tachyon with
this potential behaves as dust or dark energy.}

\vskip 0.4cm
\section{Introduction}
%{\bf 1.} 

Observations of the Cosmic Microwave Background \cite{WMAP} and of distant type Ia
supernovas \cite{SN} indicate that the universe has been accelerating its
expansion rate for the last 5-10 Gy.
Prime candidates for causing this recent burst of expansion are a
cosmological constant and a scalar field \cite{RP,accellera} which, for
reasons unknown yet, started to dominate over the other types of matter just at our 
cosmological era.
A phase of accelerated expansion driven by a scalar field (inflation) is also
the preferred theory for the early universe, since it explains the homogeneity,
flatness and the near-scale-invariant spectrum of cosmological perturbations of the 
Universe.

Most scalar models of accelerated expansion involve {\it canonical} scalar field theories:
\be
\label{L_csf}
{\cal{L}}_{C} = - \sqrt{-g} \, \left[ \frac12 \partial^\mu \varphi \partial_\mu \varphi
+ V(\varphi) \right] \; .
\ee
Another class of scalar field theory, which appeared first as a certain field theory 
generalization of the Lagrangian of a relativistic particle\cite{padma}, is the 
{\it Born-Infeld} theory:
\be
\label{L_BI}
{\cal{L}}_{BI} = - \sqrt{-g} \, V({\varphi}) \, 
\sqrt{1 + \frac{\partial^\mu \varphi \partial_\mu \varphi}{M^4}} \; .
\ee
Such a kind of scalar field was proposed in connection with
string theory, since it seems to represent a low-energy effective
theory of D-branes and open strings, and has been conjectured to play a role 
in cosmology \cite{Sen,padma}. Since the scalar field in the Lagrangian 
(\ref{L_BI}) stands for the {\it tachyon} of string theory, that name was also
attached to the present scalar field.
%, although in this incarnation the field is by no means unstable.

In the next Section we briefly review the dynamics of the tachyon scalar field.
In Section 3 we determine the asymptotic tachyon-dominated backgrounds. 
In Section 4 we consider inhomogeneous perturbations in inflating and dust tachyon 
models. In Section 5
we determine how the tachyon will behave in the presence of another
background fluid.

\section{Dynamics of the tachyon}

The equation of motion for the tachyon is:
\be
\label{eqtach}
- \nabla^\mu \partial_\mu \varphi + \frac{\nabla_\mu \partial_\nu
\varphi}{M^4 + \partial_\mu \varphi \partial^\mu \varphi} \partial^\mu
\varphi \partial^\nu \varphi + M^4 (\log V)_\varphi = 0 \,.
\ee
At the homogeneous level this becomes:
\be
\label{eom_T}
\frac{\ddot \varphi}{1 - \frac{\dot \varphi^2}{M^4}} + 3 H \dot \varphi + 
M^4 (\log V)_\varphi = 0 \; .
\label{eq_homo}
\ee
The energy and pressure densities of the tachyon are:
\begin{eqnarray} 
\label{rho_T}
\rho_T &=&  \frac{V(\varphi)}{\sqrt{1 - \frac{\dot \varphi^2}{M^4}}} \; ,\\
\label{p_T}
p_T &=& - V(\varphi) \sqrt{1 - \frac{\dot \varphi^2}{M^4}} \; .
\end{eqnarray} 
The tachyon fluid is also characterized by the ratio between pressure and 
energy (the equation of state) $w_t$ and sound speed $c_T^2$:
\be
w_T = - 1 + \frac{\dot \varphi^2}{M^4} \quad , \quad c_T^2 = - w_T  \; .
\ee
Since the equation of state is necessarily nonpositive because of the
square root in the action (\ref{L_BI}), the theory is stable
--- energy and pressure are real, and inhomogeneous perturbations
have a positive sound speed.
%are equally well behaved. 
Moreover, because $w_T \le 0$, the tachyon is a 
natural candidate for dark energy and inflation.

Another interesting property is that the
equation of state and sound speed of tachyons are equal, but with opposite signs, 
irrespective of the form for the potential. Canonical scalar fields, on the other hand,
obey the Klein-Gordon equations, hence its fluctuations travel with sound
speed equal to unity (in units where $c=1$).
Therefore, tachyon fluctuations are fundamentally different from the fluctuations in
a canonical scalar field, irrespective of the shape of the potential.

\section{Tachyon-dominated backgrounds}

First, we study the evolution of the tachyon as the only component of the 
universe -- i.e., we assume that tachyons are the dominant component.
The expansion rate is then given only by the tachyon energy density in Eq. (\ref{rho_T}):
\be
\label{H_rho_T}
%3H^2 = M_{Pl}^{-2} 
H^2 = \frac{1}{3 M_{\rm pl}^2}
\frac{V(\varphi)}{\sqrt{1-\frac{{\dot\varphi}^2}{M^4}}} \; ,
\ee
where $M_{Pl}^{-2}=8\pi G$. Defining the new variables:
\bea
\label{def_y}
y &\equiv& 1-\frac{\dot\varphi^2}{M^4} = -w_T  \; , \\
\label{def_x}
x &\equiv& V(\varphi) \; ,
\eea
we can reduce the equation of motion (\ref{eom_T}) to the form:
\be
\label{eom_y}
\frac{ d \ln y}{d x} = {\rm sign}[\dot\varphi] \, y^{-1/4} \sqrt{1-y} \, h(x) + \frac{2}{x} \; ,
\ee
where
\be
\label{h_x}
h(x) \equiv \frac{2\sqrt{3}}{M_{\rm Pl} M^2}  \frac{V^{1/2}}{V_{,\varphi}} 
\; .
\ee

Let us assume now that the tachyon potential is an inverse power-law of the tachyon field:
\be
\label{pot_T}
V = \frac{ m^{4+\alpha}}{\varphi^{\alpha}} \; ,
\ee
with $\alpha > 0$ and also $\dot\varphi > 0$ (exotic initial conditions 
could be concocted to produce
$\dot\varphi < 0$ with these potentials, but they are irrelevant to the asymptotic 
behaviour of these models.)
The asymptotic behaviour of this type of potential is in agreement with the spirit of the
original string-inspired proposal \cite{Sen}, in which the 
rolling tachyon field describes the low-energy sector for D-branes 
and open strings. Its potential should go to zero at infinite values of 
the field, such that in this asymptotic
vacuum there would be no D-branes, and thus no open strings.
Originally the potential was determined to be exponential
%:$V \sim \exp{[ - \kappa \varphi ]}$ 
\cite{Sen}.
The exponential potential can be obtained from the power-law ansatz 
for the potential 
by taking appropriately the limit $\alpha \rightarrow \infty$.

Assuming a potential of the form (\ref{pot_T}) we have:
\be
\label{h_pl}
h(x)=-\frac{2\sqrt{3}}{\alpha} M_{Pl}^{-1} M^{-2} m^{\frac{4+\alpha}{\alpha}} x^{-\frac{2+\alpha}{2\alpha}} \; ,
\ee
which allows us to rewrite equation (\ref{eom_y}) as:
\be
\label{eom_y_s}
\frac{ d \ln y}{d x} = - \beta y^{-1/4} \sqrt{1-y} \, 
x^{-\frac{2+\alpha}{2\alpha}} + \frac{2}{x} \; ,
\ee
where the constant $\beta=2 \sqrt{3}  \alpha^{-1} M_{Pl}^{-1} M^{-2} m^{(4+\alpha)/\alpha}$.

For $\alpha=2$ this equation admits an exact solution with $y=$constant
\cite{padma,feinstein}. This corresponds to a power-law evolution 
of the scale factor $a(t) = a_0 t^p$ with:
\begin{eqnarray}
p &=& \frac{1}{3} \left(1 + \sqrt{1 + \frac{9}{4} \frac{m^{12}}{M^8 M_{\rm pl}^4}} \; \right) \; ,
\\
\varphi (t) &=& \sqrt{\frac{2}{3 p}} M^2 t \; .
\end{eqnarray}

For $\alpha \neq 2$ the system (\ref{eom_y_s}) has an attractor as $x \rightarrow 0$
either at $y=0$ (dust) or at $y=1$ (quasi-de Sitter), respectively when $\alpha$ is greater
or smaller than 2.

The dust solution close to the point $(y \rightarrow 0,x \rightarrow 0)$ can be calculated from
Eq. (\ref{eom_y_s}) by assuming the term inside the square root to be close to unity. 
We then have the approximate solution when $x \rightarrow 0$:
\be
\label{sol_dust}
y \simeq \gamma x^s \; ,
\ee
where
\bea
\label{s}
s &=& \frac{2(\alpha-2)}{\alpha} \; , \\
\label{gamma}
\gamma &=& \frac{9}{16} \frac{m^{\frac{4(4+\alpha)}{\alpha}}}{M_{Pl}^4M^8} \; .
\eea
The condition under which this solution satisfies the assumption that $y \rightarrow 0$ as
$x \rightarrow 0$ is that $\alpha > 2$. Therefore, for powers of the tachyon potential which are
bigger than 2, the late-time behaviour of the system is that 
of a dust-dominated universe \footnote{This result was independently found 
in \cite{felder}. We thank Prof. Starobinsky for pointing this out to us.}.
In particular, for $\alpha \rightarrow \infty$ the potential becomes 
exponential and one obtains
that the dust solution is approached exponentially fast \cite{frolov}.

We now prove that this dust-like solution is stable.
A small fluctuation $\delta y(x)$ around the solution (\ref{sol_dust}), using 
Eq. (\ref{eom_y_s}), obeys:
\be
\label{fluc_y_d}
\frac{d \ln \delta y}{dx} = \frac{2\alpha-3}{\alpha}  \, \frac{1}{x} \; ,
\ee
and therefore the deviation from the dust solution decays as
\be
\label{dec_y_d}
\frac{\delta y}{y} \sim x^{1/\alpha} \; .
\ee

The quasi-de Sitter solution, for which $y \rightarrow 1$ as $x \rightarrow 0$, can be obtained 
by matching the terms in the R.H.S. of Eq. (\ref{eom_y_s}), and then improving this
solution. The improved solution is:
\be
\label{sol_z}
y \simeq 1 - \frac{M_{Pl}^2 M^4}{3 m^6} \; \left( \frac{x}{m^4} \right)^{-1+2/\alpha} 
\left[ 1 - \frac{3\alpha-4}{\alpha} \frac{M_{Pl}^2 M^4}{12 m^6} 
\; \left( \frac{x}{m^4} \right)^{-1+2/\alpha} \right] \; .
\ee
The first term decays when $\alpha < 2$, which means that this solution approaches
de Sitter as $x \rightarrow 0$.
The last term in the solution above is a further correction that 
decays even faster. Therefore, the quasi-de Sitter solution is also stable against small 
fluctuations.

Since the type of acceleration one obtains is perhaps of a peculiar sort,
we explicitly derive it. From Eq. (\ref{sol_z}) one obtains that 
$\dot \varphi \propto \varphi^{1-2/\alpha}$, and substituting this into Eq. (\ref{H_rho_T})
one obtains, for the leading term:
\be
\label{H_qdS}
H = H_0 \left( \frac{t}{t_0} \right)^{-\frac{\alpha}{4-\alpha}} \; ,
\ee
and therefore
\be
\label{a_qdS}
a(t) = a_0 \exp{\left[ \left( \frac{t}{t_0} \right)^{\frac{4-2\alpha}{4-\alpha}} \right]} \; .
\ee
For $\alpha \rightarrow 0$ one obtains exact de Sitter with $3 H^2 = m^4/M_{pl}^2 $. 
Indeed, the limit $\alpha \rightarrow 0$ corresponds to the so-called Chaplygin Gas 
\cite{Chap},
which develops into de Sitter after a period of dust-like behaviour obtained by appropriately
handling initial conditions.
For $\alpha \rightarrow 2$ one obtains from (\ref{a_qdS})
a power-law expansion, in agreement with the first case discussed in this Section.

Therefore, for inverse power-law tachyon potentials $V\sim \varphi^{-\alpha}$, the asymptotic
behaviour is very simple: a quasi de Sitter spacetime if $0 < \alpha < 2$, a power-law expansion
if $\alpha=2$, or a dust-like Universe if $\alpha > 2$. 
The asymptotic tachyon dynamics is summarized in Fig. 1.

As already emphasized, a component with non-positive pressure is
interesting for inflation. Inflation driven by a tachyon field has been
promptly criticized \cite{kofmanlinde} since the effective 4D field theory 
derived from string theory would lead to a too high scale of
inflation (related to $m$), and thus produce an unacceptable background of
stochastic gravitational waves. Hopefully future developments of string 
theory will help resolve this issue.

%%%%%%
% HERE INCLUDE THE PHASE DIAGRAM AS AN EPS FIGURE
%%%%%%
%\begin{figure*}[t]
%  \centering
% \include{diagr_fase.eps}
%  \caption{Phase diagram}
%  \label{fig:1}
%\end{figure*}

\begin{figure} 
%\noindent
%\epsfig{file=diagr_fase.eps,height=7.0cm}
\vspace{4.5cm}
\includegraphics{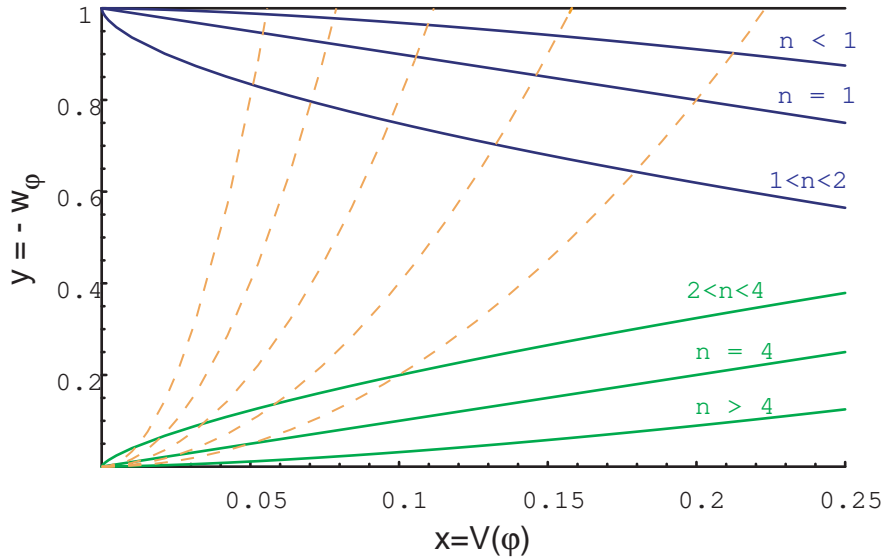}
\caption{Phase diagram for rolling tachyons with the inverse power-law 
potential $V(\varphi)=m^{4+\alpha} \varphi^{-\alpha}$. The dashed lines 
are constant energy density surfaces.}
\label{fig1}
\end{figure}

For negative values of $\alpha$, the fate of the tachyon is to oscillate around
the minimum of the potential with an averaged equation of state which is negative
\cite{frolov}. That the tachyon energy density $\rho_T = x/\sqrt{y}$ is well behaved at the 
minimum of the potential can be seen from considering the solution $y \propto x^2$
which is found by neglecting the first term in the R.H.S. of
Eq. (\ref{eom_y_s}) as $x\rightarrow 0$. But if $y\propto x^2$, then the neglected
term goes as $x^{-1-1/\alpha}$, which means that it is subdominant with respect to the
second term when $\alpha<0$.
However, notice that by Eq. (\ref{eqtach}) small fluctuations of the tachyon field
have an effective mass squared equal to:
\be
\label{effmass}
m_{eff}^2 = M^4 \left( \ln V \right)_{,\varphi \varphi} = \alpha \frac{M^4}{\varphi^2} \; .
\ee
This means that for negative values of $\alpha$ the tachyon field is unstable under
small perturbations \cite{frolov}. Here we focus on {\it positive} values of $\alpha$,
for which the fluctuations have a positive effective mass.

Since the tachyon 
energy density decays always slower than that of dust, in the absence of a cosmological 
constant it is bound to eventually dominate over the other components (matter and radiation),
and it is therefore a natural candidate for inflation and dark energy.
Notice that tachyon fluctuations have a sound speed equal to $c_T^2=-w_T$, and hence
have an increased tendency to cluster on sub-Hubble scales compared to standard quintessence 
\cite{usprep}.
However, if it is to serve as a viable candidate for dark energy, it must have been
subdominant in the past. In the section 5 we analyse the tachyon dynamics 
in the
presence of a perfect fluid-dominated background.

\section{Cosmological Perturbations}

We discuss gravitational fluctuations for the case in which tachyon is the
only component of the universe. Gravitational waves just feel
the evolution of the scale factor and in the long-wavelength limit
the tensor amplitude $h_k$ behave as:
\be
h_k = A_k + B_k \int \frac{d t}{a^3}
\ee
where $A_k$ and $B_k$ are functions of $k$. Besides the constant mode
$A_k$, the solution containing $B_k$ is decaying for the type of
inflationary and dust solutions discussed in this paper.

Scalar metric fluctuations
are described by the Mukhanov variable $v$ whose evolution is \cite{GM}
\be
v_k'' + ( c_T^2 k^2 - \frac{z''}{z} ) v_k = 0 \,,
\ee
where ' denotes derivative with respect to the conformal time ($d \eta =
d t/a$), and \cite{frolov} :
\be
z = a \frac{\sqrt{\rho_T+p_T}}{c_T H} = a \sqrt{3}
\frac{M_{\rm pl}}{M^2} \frac{\dot \varphi}{\sqrt{1 - \frac{\dot
\varphi^2}{M^4}}} \; .
\ee
The solution for $v_k$ in the long-wavelength limit is:
\be
v_k = C_k z + D_k z \int \frac{d t}{a z^2} \,,
\ee
where $C_k$ and $D_k$ are functions of $k$.
The curvature perturbation $\zeta \equiv v/z$ in the long-wavelength
limit is therefore:
\be
\zeta_k = C_k + D_k \int \frac{d t}{a z^2} \,.
\ee
Besides the constant mode, the solution containing $D_k$ is decaying for
the type of inflationary and dust solutions discussed in this paper.

The curvature perturbation $\zeta$ remains constant on large scales, even
if the gauge invariant field fluctuation $\delta \varphi = v /a$ can vary
in time:
\be
\delta \varphi_k \propto \frac{\dot \varphi}{c_T} \,.
\ee
More precisely, on large scales, tachyon fluctuations decrease during
the quasi de Sitter regime and grow in the dust regime. For the threshold 
$\alpha = 2$ case long wavelength tachyon fluctuations are constant in 
time. 

In the dust regime metric perturbations
remain constant on large scales. In fact, even the Newtonian potential 
$\Phi$, which
satisfy:
\be
\zeta = \frac{2}{3} \frac{\dot \Phi/H + \Phi}{1 + w_T} + \Phi
\ee
remains constant on long-wavelength scales, satisfying the usual relation
$\Phi \simeq 3/5 \zeta$ of matter-dominated era. However, long-wavelength field 
fluctuations build up in time as:
\be
\delta \varphi \sim t^{\alpha-1} \rightarrow \delta \varphi/\varphi \sim
t^{\alpha-2}
\ee
leading to a non-linear stage for the tachyon field.
Metric fluctuations are however kept constant by a compensating term.
These considerations hold when the tachyon behaves as dust,
irrespective of the form of the potential. At the level of linear
perturbation theory the metric fluctuations
stay constant even as the field fluctuations grow on large scales: it 
has been argued that also metric fluctuations at large scales will grow at 
the nonlinear level \cite{felder}.

%The stage of non-linearity for a tachyon in the dust regime is not just
%due to the field representation. In fact, even by considering the fluid
%representation, one finds that the velocity potential $\theta$, defined
%as:

We would also like to emphasize that the effective mass for tachyon 
fluctuations in Eq. (\ref{effmass}) is really different from the scale of inflation. This is 
attractive if one thinks at the criticism about the flatness of the 
potential \cite{afg}. In the spirit of k-inflation \cite{GM}, the prediction about the
ratio of scalar to tensor perturbations will be different from the one of canonical
scalar field models. So, in principle inflation with the tachyon can be distinguished 
from canonical scalar field inflation.

\section{Tachyon and Perfect Fluid Dynamics}

We now study the evolution of the tachyon fluid in the background of a 
perfect fluid characterized by energy density $\rho_F$ and constant 
equation of state $w_F$. The Hubble law is therefore:
\be
H^2 = \frac{1}{3 M_{\rm pl}^2} \left[ \frac{V(\varphi)}{\sqrt{1 - 
\frac{\dot \varphi^2}{M^4}}} + \rho_F \right]
\label{hubble}
\ee
The equation of motion for the tachyon (\ref{eq_homo}) can be satisfied for 
$\dot \varphi =$constant. This corresponds to a perfect fluid-like evolution 
for the tachyon energy density. The particular case 
$\dot\varphi \sim M^2$ ($w_T \sim 0$) 
is interesting for the tachyon as a candidate for CDM.
However, other fluid-like attractors exist for the tachyon with a negative 
power-law potential (\ref{pot_T}) in the presence of another dominating fluid if 
$0 < \alpha < 2$.

If the tachyon component is subdominant with respect to the background fluid, 
Eq. (\ref{hubble}) means that:
\be
H \simeq \frac{2}{3 (1 + w_F) t} \,.
\ee
With the potential (\ref{pot_T})
a solution for the homogeneous tachyon field in which its amplitude 
grows linearly in time is possible:
\be
\varphi \simeq A \, t \quad {\rm with} \quad A = M^2 
\sqrt{\alpha \frac{1+w_F}{2}} \,,
\label{tracking}
\ee
where we have neglected the initial value $\varphi_0$ for the tachyon
and a negative sign for the tachyon velocity (a tachyon climbing up
its effective potential.)
Neglecting the initial value translates in an inequality for the 
growing tachyon:
\be
\varphi_0 H_0 << M^2 \, \sqrt{\frac{2 \alpha}{9 (1 + w_F)}} 
\ee
%and in this equality for the decreasing tachyon:
%\be
%\varphi_0 H_0 = M^2 \, \sqrt{\frac{2 \alpha}{9 (1 + w_F)}}
%\ee
where the subscript $0$ denotes some primordial time. 

Small deviations $\delta y_A (t)$ around the tracking solution 
(\ref{tracking}) satisfy the following equation:
\be
\delta \ddot y_A + \left( \frac{2}{1+w_F} - \alpha \right) \frac{\delta \dot y_A}{t} + 
\left( \frac{2}{1+w_F} - \alpha \right) \frac{\delta y_A}{t^2} = 0 \,.
\ee 
The solutions are of power-law type, i.e. $\delta y_A \sim t^{\gamma_A}$, with:
\be
\gamma_A = \frac12 (1+\lambda) \pm \frac12 \sqrt{(1+\lambda)^2+4\lambda} \; \; ,
\ee
where $\lambda = \alpha - 2/(1+w_F)$. This
shows that the tracking solutions (\ref{tracking}) are stable for 
$\lambda < 0$, i.e. $\alpha< 2/(1+w_F)$.

For this class of solution
the tachyon field has an equation of state:
\be
w_T = - 1 + \frac{A^2}{M^4} = - 1 + \alpha \frac{1+w_F}{2} \,.
\ee
However, these solutions can only exist if they respect the condition that the tachyon
equation of state is nonpositive. This means that, for instance, the fluid-like 
solution is only valid in the range $0<\alpha<3/2$ if the background is radiation 
($w_F=1/3$) \footnote{In \cite{chiba} it was argued 
$0 < \alpha < 2$ for the general case 
in which the Lagrangian is separable: ${\cal L} = p (\varphi, X) = V 
(\phi) W (X)$, where $X=\partial^\mu \varphi \partial_\mu \varphi$. 
We obtain a different value because we do not assume $w_F = 0$ (dust) for
the tracking.}.

For the range $2/(1+w_F) < \alpha < 2$ the attractor saturates the dust 
limit --- i.e., $w_T \rightarrow 0^-$. The solution of Eq. (\ref{eq_homo}) in that 
case is given by:
\be
\label{dust_gen}
\varphi = M^2 t - \epsilon t^{1-2 \left( \alpha - \frac{2}{1+w_F} \right) } \; ,
\ee
where $\epsilon$ is an arbitrary constant that depends on the initial 
conditions. The correction term decays in time compared to the leading term
if $\alpha > 2/(1+w_F)$, showing
that dust is indeed an attractor for this range of parameters.

A curious property of tachyons as candidates for dark energy is that when they
are in a background which is dominated by a fluid with equation of state
$w_F$, their energy density behaves either as $\rho_T \propto a^{-3}$ when 
$\alpha > 2/(1+w_F)$, 
or as $\rho_T \propto t^{-\alpha}$ if $0<\alpha<2/(1+w_F)$.
But on the other hand, when the tachyon starts to dominate, it will behave
as dust if $\alpha > 2$ or it will drive quasi-de Sitter acceleration if
$0<\alpha<2$. Therefore, if $2/(1+w_F) < \alpha < 2$, a change in the
equation of state of the background (as, for example, the change from radiation-
to matter-domination) will trigger a transmutation in the tachyon, causing it to
leave the dust attractor and start to accelerate the expansion rate (a similar
behavior was found in the ``k-essence'' model \cite{kessence}).

\section{Conclusions}

We have analysed the cosmological dynamics of a Born-Infeld
scalar field (also know as the tachyon) with an inverse power-law potential. 
We found that the tachyon has a very simple behavior: if the tachyon dominates
the background dynamics, then it will either go into a dust-dominated phase
($\alpha > 2$), power-law expansion with a constant $w_T < 0$ for 
$\alpha=2$, or
quasi-de Sitter accelerated expansion for $0<\alpha<2$. While during a 
dust stage tachyon fluctuations are driven to a non-linear 
stage, the linear evolution of tachyon fluctuations is valid in the other 
cases.

%\begin{figure}
%\epsfxsize=2 in \epsfbox{h070_X70_A=100B.ps}
%\caption{}
%\label{fig1}
%\end{figure}

%\vspace{.1cm}

\vskip 1cm

{\bf Acknowledgments}

\noindent
We would like to thank A. Starobinsky for comments and I. Waga for useful 
conversations.
F. F. would like to thank the Instituto de F\'{\i}sica, 
Universidade de S\~ao Paulo, for its warm hospitality when
this work was initiated. R. A. would like to thank the CNR/IASF - Bologna
as well, for its hospitality. This work was supported by FAPESP and 
CNPq (Brazil), and by the CNR/IASF - Bologna (Italy).

%\section{Appendix}

%\vspace{.1cm}

\end{document}